\documentclass[english,aps,prl,amsmath,amssymb,twocolumn,letterpaper,showpacs,superscriptaddress]{revtex4-1}
\usepackage[latin9]{inputenc}
\usepackage{babel}
\usepackage{verbatim}
\usepackage{amsmath}
\usepackage{amssymb}
\usepackage[unicode=true,pdfusetitle,
 bookmarks=true,bookmarksnumbered=false,bookmarksopen=false,
 breaklinks=false,pdfborder={0 0 1},backref=false,colorlinks=false]
 {hyperref}

\usepackage{eso-pic} 
\usepackage{graphicx}

\usepackage{pdfpages}	
\usepackage{pgffor}		

\makeatletter
\AtBeginDocument{\let\LS@rot\@undefined} 
\makeatother							 

\makeatother

\begin{document}
\title{Superconducting correlations out of repulsive interactions on a fractional
quantum Hall edge}
\author{Jukka I. V\"ayrynen}
\affiliation{Microsoft Quantum, Microsoft Station Q, University of California,
Santa Barbara, California 93106-6105 USA}
\author{Moshe Goldstein}
\affiliation{Raymond and Beverly Sackler School of Physics and Astronomy, Tel Aviv
University, Tel Aviv 6997801, Israel}
\author{Yuval Gefen}
\affiliation{Department of Condensed Matter Physics, Weizmann Institute of Science,
Rehovot 76100, Israel}
\date{\today}
\begin{abstract}
 We consider a fractional quantum Hall bilayer system with an interface
between quantum Hall states of filling fractions $(\nu_{\text{top}},\nu_{\text{bottom}})=(1,1)$
and $(1/3,2)$, motivated by a recent approach to engineering artificial
edges~\cite{2018NatPh..14..411R}. We show that random tunneling
and strong repulsive interactions within one of the layers will drive
the system to a stable fixed point with two counterpropagating charge
modes which have attractive interactions. As a result, slowly decaying
correlations on the edge become predominantly superconducting.  We
discuss the resulting observable effects, and derive general requirements
for electron attraction in Abelian quantum Hall states. The broader
interest in fractional quantum Hall edge with quasi-long range superconducting
order lies in the prospects of hosting exotic anyonic boundary excitations,
that may serve as a platform for topological quantum computation.
\end{abstract}
\maketitle
\textbf{\emph{Introduction. }}Combining superconductivity and fractional
quantum Hall edge states opens the possibility to engineer exotic
topological phases of matter with anyonic boundary excitations~\citep{PhysRevX.2.041002,2013NatCo...4E1348C,2012PhRvB..86s5126C,2013PhRvB..87c5132V,PhysRevX.4.011036,2018PhRvL.120f6801H,alicea2016topological}.
A possible route to this is by using the proximity effect with a bulk
superconductor and a quantum well in a hybrid structure~\citep{2016Sci...352..966A,2017NatPh..13..693L,2016NatPh..12..318B}.
Another, less studied possibility is that of intrinsic superconductivity
on the edge. Evidently, on a one-dimensional edge there is no true
long-range order and correlation functions decay algebraically. One
can nevertheless refer to a superconducting phase as the one where
the slowest decaying correlation function is of superconducting nature,
i.e., a pairing correlator~\citep{1979AdPhy..28..201S}. Such power-law
(or quasi-long-range) superconducting order may still be relevant
for topological quantum computing applications~\citep{2008RvMP...80.1083N},
c.f.~\citep{PhysRevB.84.195436} in the context of Majorana bound
states. \emph{}

In a recent experimental work~\cite{2018NatPh..14..411R} it has
been demonstrated that in an engineered bilayer system it is possible
to structure and control co- and counterpropagating edge modes in
both the integer and fractional quantum Hall regimes. The present
work takes advantage of this new paradigm and shows that one can design
chiral modes with bare repulsive interaction in the presence of disorder
to induce \emph{attractive} interaction between the resulting effective
modes. This gives rise to a phase with algebraically-decaying superconducting
order.

\begin{figure} \includegraphics[width=1\columnwidth]{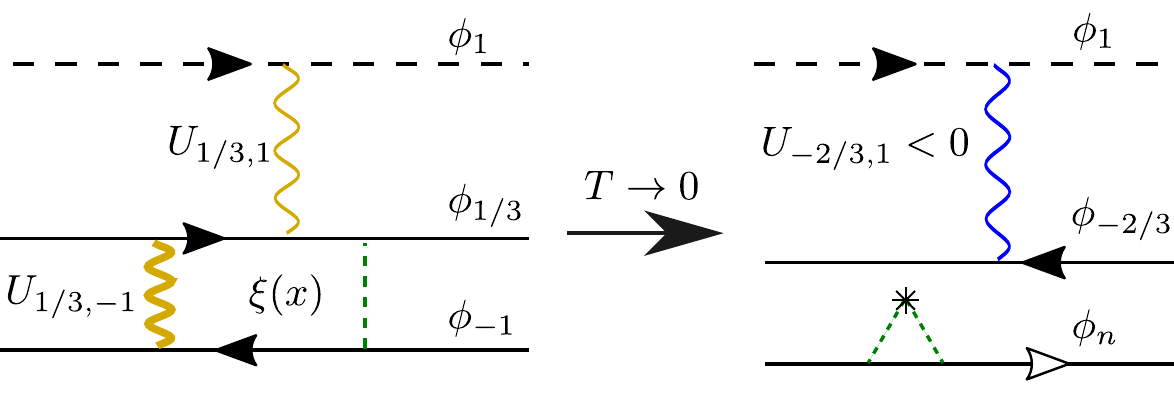}\caption{ Pictorial description of the main result. On the left, we show the configuration of the bare edge modes. 
The configuration of edge modes can be experimentally realized in a bilayer structure, see Fig.~\ref{fig:2}a. There is strong Coulomb interaction, $U_{1/3,-1}$, as well as random tunneling, $\xi(x)$, between the modes $\phi_{1/3}, \,\phi_{-1}$. The third mode $\phi_{-1}$ is weakly coupled with repulsive interaction to the mode $\phi_{1/3}$. As temperature is lowered the random tunneling renormalizes the strongly-coupled modes to a fixed point described by a neutral mode $\phi_{n}$ and a charge mode $\phi_{-2/3}$, shown on the right. Remarkably, the interaction between the charge modes  $\phi_{-2/3}$ and $\phi_{1}$ can be attractive while the neutral mode is decoupled due to the disorder.  
   \label{fig:1}} \end{figure}

To describe our results qualitatively, let us recall the pioneering
work~\citep{kane_randomness_1994} of Kane, Fisher, and Polchinski
(KFP) for $\nu=2/3$ edge hosting counterpropagating $\nu=1/3$ and
$\nu=1$ modes. Random tunneling and sufficiently strong interaction
between the two modes can drive the system to a fixed point with decoupled
neutral and charge modes. The charge of the latter, $2e/3$, is determined
by the constituent bare modes and charge conservation. The fixed point
is approached upon, for example, lowering the temperature, and can
be understood as a renormalization of the interaction between the
neutral and charge mode (the interaction is an irrelevant perturbation
and renormalizes to zero). The novel aspect in our proposal is to
consider an additional $\nu=1$ ($1e$) charge mode interacting with
the KFP  modes, see Fig.~\ref{fig:1}. As in the conventional KFP
theory, both charge modes decouple from the neutral mode upon decreasing
temperature. However,  now there is a set of KFP fixed points, parametrized
by the interaction between the $2e/3$ and $1e$ charge modes. Our
main finding is that this fixed-point interaction can be \emph{attractive},
even when the bare interactions of the high-temperature limit are
repulsive.  We further substantiate this claim by studying the renormalization
group flow in a fine-tuned strongly-interacting model where the $2e/3$
and the neutral mode are already decoupled on the level of the bare
Hamiltonian. We then move on to study the new fixed point. We find
that the fixed point has superconducting correlations of the charge
modes: their pairing correlation function decays slower than any charge
density correlation function. Finally, we outline how our model can
be realized in an engineered quantum Hall bilayer system and how one
can detect the attractive interactions at the fixed point by using
3 experimental probes: multiterminal shot noise, tunneling spectroscopy,
and ground state charge in a quantum dot geometry.

\textbf{\emph{Model. }}\emph{}%
{} We consider a system with 3 relevant edge modes. We assume a right-moving
$\nu=1/3$ mode and a pair of  counterpropagating $\nu=1$ modes.
In terms of a three-component chiral boson field $\boldsymbol{\phi}=(\begin{array}{ccc}
\phi_{1/3} & \phi_{-1} & \phi_{1}\end{array})^{T}$, our model is described by the imaginary-time action 
\begin{flalign}
S & =\int d\tau dx\frac{1}{4\pi}\left[\partial_{x}\boldsymbol{\phi}\mathbf{K}i\partial_{\tau}\boldsymbol{\phi}+\partial_{x}\boldsymbol{\phi}\mathbf{V}\partial_{x}\boldsymbol{\phi}\right]\label{eq:ShomogDiab}\\
 & +\int d\tau dx\left[\xi(x)e^{i\mathbf{c}\cdot\boldsymbol{\phi}}+\xi^{*}(x)e^{-i\mathbf{c}\cdot\boldsymbol{\phi}}\right]\,,\nonumber 
\end{flalign}
where in the first line $\mathbf{K}=\text{diag}(3,-1,1)$  and the
$V$-matrix is 
\begin{equation}
\mathbf{V}=\left(\begin{array}{ccc}
3v_{1/3} & U_{1/3,-1} & U_{1/3,1}\\
U_{1/3,-1} & v_{-1} & U_{-1,1}\\
U_{1/3,1} & U_{-1,1} & v_{1}
\end{array}\right)\,.\label{eq:V}
\end{equation}
We assume that the mode $\phi_{1}$ is physically far from the other
two, so that $U_{1/3,1},\,U_{-1,1}\ll U_{1/3,-1}$. Finally, the second
line of Eq.~(\ref{eq:ShomogDiab}) describes disordered tunneling
of electrons between the counterpropagating $1/3$ and $1$ modes;
here the tunneling vector is $\mathbf{c}=(3,1,0)$ and $\xi$ is
a $\delta$-correlated random coefficient, $\left\langle \xi(x)\xi(x')\right\rangle =W\delta(x-x')$,
with zero average. The random tunneling term is non-linear in the
boson fields and leads to non-trivial renormalization of the $V$-matrix. 

\emph{}%

Let us first ignore the interactions $U_{1/3,1},\,U_{-1,1}$ and consider
the problem of two modes $\phi_{1/3},\:\phi_{-1}$. This is exactly
the model studied by KFP~\citep{kane_randomness_1994} in the context
of the edge of the $\nu=2/3$ quantum Hall state. The amplitude of
random tunneling obeys the renormalization group (RG) equation~\citep{giamarchi_anderson_1988}
$\frac{dW}{dl}=(3-2\Delta_{3,1,0})W$. Here the scaling dimension
is  $\Delta_{3,1,0}=(2-\sqrt{3}c)/\sqrt{1-c^{2}}$ where $c=(2U_{1/3,-1}/\sqrt{3})/(v_{1/3}+v_{-1})$;
the perturbation is relevant, $\Delta_{3,1,0}<3/2$, when $0.34\lesssim c\lesssim0.98$.
Thus, for sufficiently large positive (repulsive interaction) $U_{1/3,-1}$,
the random tunneling operator $e^{i[3\phi_{1/3}+\phi_{-1}]}$ is a
relevant perturbation and its amplitude grows upon lowering the temperature.
This tunneling operator is the only relevant one as long as we ignore
tunneling to the mode $\phi_{1}$. The latter mode can be ignored
due to its larger separation~\cite{2018NatPh..14..411R}, see also
Discussion below. Next, we study how the increasing $W$ under RG
transformation affects the elements of the $V$-matrix. 

\textbf{\emph{Neutral mode basis and perturbative RG. }}\emph{}%
Following Ref.~\citep{kane_randomness_1994}, it is convenient to
work in the basis where the random tunneling $e^{i[3\phi_{1/3}+\phi_{-1}]}$
is diagonal. This is the basis of a right-moving neutral mode and
a left-moving charge mode, 
\begin{equation}
\phi_{n}=\frac{1}{\sqrt{2}}\left(3\phi_{1/3}+\phi_{-1}\right)\,,\quad\phi_{-2/3}=\sqrt{\frac{3}{2}}\left(\phi_{1/3}+\phi_{-1}\right)\,.\label{eq:neutralcharge}
\end{equation}
The random tunneling, which conserves charge, only couples to $\phi_{n}$:
in this ``neutral mode basis'' $\boldsymbol{\phi}=(\begin{array}{ccc}
\phi_{n} & \phi_{-2/3} & \phi_{1}\end{array})_{(n)}^{T}$ and the tunneling vector becomes $\mathbf{c}=(\sqrt{2},0,0)_{(n)}$.
Also, $\mathbf{K}=\text{diag}(1,-1,1)_{(n)}$ and 
\begin{equation}
\mathbf{V}=\left(\begin{array}{ccc}
v_{n} & U_{n,-2/3} & U_{n1}\\
U_{n,-2/3} & v_{2/3} & U_{-2/3,1}\\
U_{n1} & U_{-2/3,1} & v_{1}
\end{array}\right)_{(n)}\,,\label{eq:Vn}
\end{equation}
where the matrix elements are simple linear combinations of the elements
from Eq.~(\ref{eq:V}). In particular, the interaction between the
charge modes $\phi_{-2/3},\,\phi_{1}$ is  $U_{-2/3,1}=\frac{1}{\sqrt{6}}(3U_{-1,1}-U_{1/3,1})$.
We see that the interaction is attractive, $U_{-2/3,1}<0$, when $U_{1/3,1}>3U_{-1,1}$.
This can happen when the mode $\phi_{1/3}$ is the nearest one to
$\phi_{1}$, as in Fig.~\ref{fig:1}. As we show below, the attractive
interaction between two charge modes makes the superconducting pair
correlations between them the slowest decaying correlation function
in the system, which we call superconducting state in 1D.

\begin{figure}[t] \includegraphics[width=0.95\columnwidth]{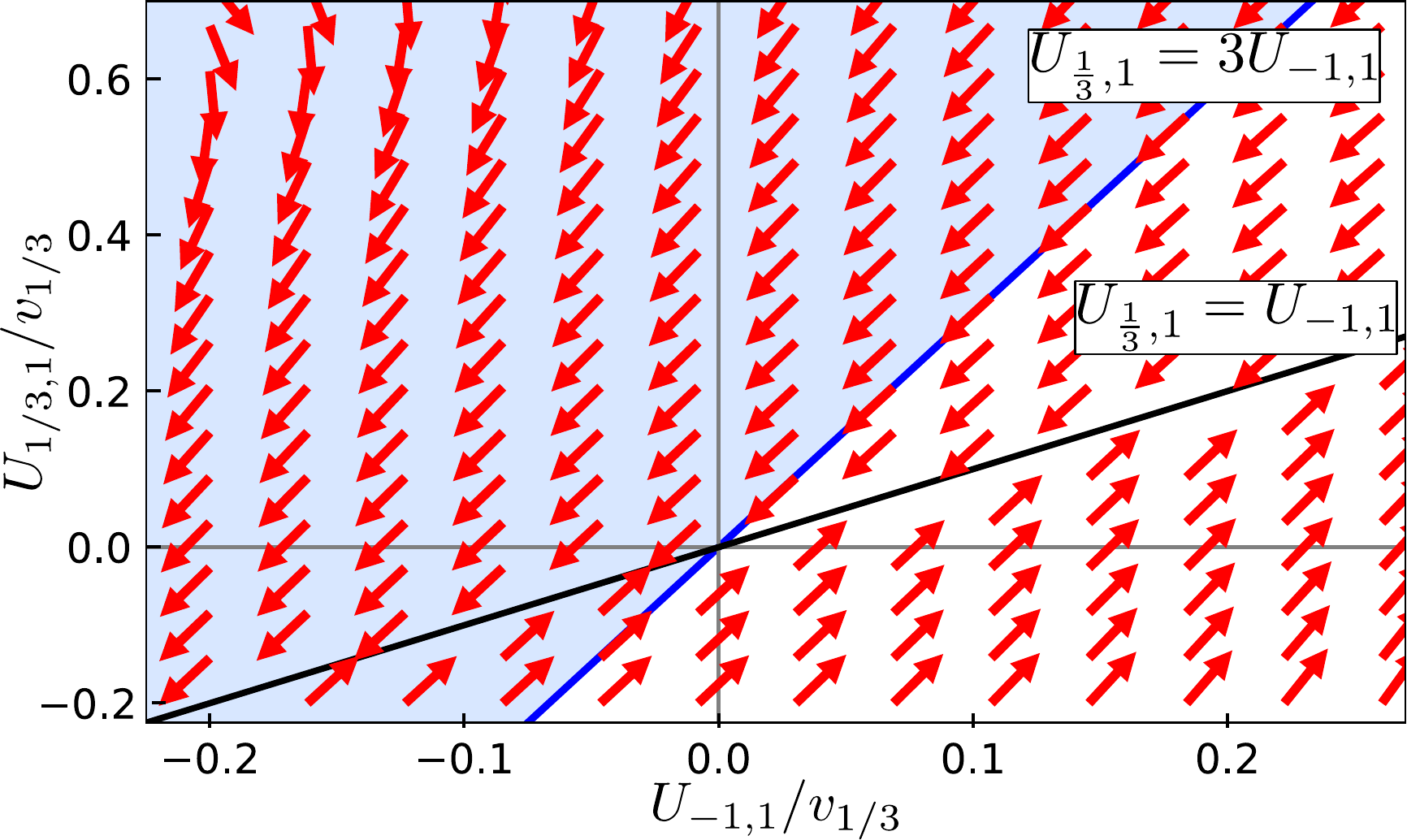}\caption{ 
Flow of the interaction parameters $U_{-1,1}\,,U_{1/3,1}$ that couple the mode $\phi_1$ to the other modes.
The bare interactions in the blue region, $3U_{-1,1}-U_{1/3,1} <0$, flow to a superconducting phase. 
There is a  line of stable fixed points $U_{-1,1}=U_{1/3,1}$ [$U_{n1} = 0$] (black). 
Near this line, the tree-level RG is accurate and flow is 
along lines of constant $3U_{-1,1}-U_{1/3,1}$ (dark blue line). 
Far from the line $U_{-1,1}=U_{1/3,1}$ one needs to account for terms beyond tree-level RG. The operator $3U_{-1,1}-U_{1/3,1}$ is   marginally irrelevant and  no longer remains constant during the flow. 
The flow diagram is calculated at the fine-tuned point where $U_{n,-2/3}=0$.
We took $v_1 = v_{-1} = 2 v_{1/3}$.
\label{fig:1b}} \end{figure}

Evidently, in the bare non-renormalized $V$-matrix the seemingly
attractive interaction is just a result of a basis change from a system
with purely repulsive interactions. The off-diagonal elements $U_{n,-2/3},\,U_{n1}$
that couple the neutral mode to the two charge modes ensure that there
are no superconducting correlations. However, we will show next that
under renormalization,  the elements $U_{n,-2/3},\,U_{n1}$ will
flow to zero due to disorder in the neutral mode, while $U_{-2/3,1}$
remains approximately constant. In the original basis this corresponds
to $U_{1/3,1},\,U_{-1,1}$ flowing to negative values, i.e., attraction,
see Fig.~\ref{fig:1b}. 

\emph{}

The weak-disorder RG flow of $\mathbf{V}$ was studied by Moore
\&~Wen~\citep{PhysRevB.66.115305}, who found that a relevant disorder
operator $e^{i\sqrt{2}\phi_{n}}$ drives the $V$-matrix towards a
fixed point which is diagonal in the neutral sector. Therefore, $U_{n,-2/3}$
and $U_{n1}$ are both irrelevant and flow to weak coupling~\footnote{The modes $\phi_{n}$ and $\phi_{1}$ are mutually chiral and therefore
a stable fixed point requires $v_{1}>v_{n}$~\citep{PhysRevB.66.115305}.}.
 Furthermore, the disorder operator $e^{i\sqrt{2}\phi_{n}}$ commutes
with $\partial_{x}\phi_{-2/3}\partial_{x}\phi_{1}$, so we expect $U_{-2/3,1}$ to be marginal, with weak renormalization stemming from its non-commutation with $U_{n i} \partial_{x}\phi_{n}\partial_{x}\phi_{i}$ ($i=1,-2/3$). 
We confirm this intuition  by finding the flow equations~\footnote{See Supplementary Material, where we present the full RG equations at $U_{n,-2/3}=0$, discuss in more detail the signatures of attraction, and outline the geometrical requirements to find attraction from repulsion in a bilayer system.}  in the limit of weak disorder and weak couplings in the KFP  fine-tuned (yet generic in terms of the resulting physics) point $U_{n,-2/3}=0$ [corresponding to $U_{1/3,-1}=3(v_{1/3}+v_{-1})/4$]. 
Numerical solution of the RG equations produces the flow diagram shown in Fig.~\ref{fig:1b},
presented in terms of the original couplings $\ensuremath{U_{-1,1}\,,U_{1/3,1}}$. 

\textbf{\emph{Strong-disorder fixed point. }}\emph{}Perturbative
treatment of random tunneling is only valid at high energies. To describe
the non-perturbative low energy regime we follow KFP and postulate
a strong-disorder fixed point $V$-matrix, 
\begin{equation}
\mathbf{V}_{\text{f.p.}}=\left(\begin{array}{ccc}
v_{n} & 0 & 0\\
0 & v_{2/3} & U_{-2/3,1}\\
0 & U_{-2/3,1} & v_{1}
\end{array}\right)_{(n)}\,.\label{eq:Vfp}
\end{equation}
At the fixed point we have a decoupled right-moving neutral mode $\phi_{n}$,
a right-moving $\nu=1$ charge mode $\phi_{1}$, and a left-moving
charge mode $\phi_{-2/3}$. The latter two are coupled via an interaction
that is attractive, $U_{-2/3,1}<0$, as long as the bare interactions
satisfy $3U_{-1,1}<U_{1/3,1}$. The set of fixed point $V$-matrices~(\ref{eq:Vfp})
can also be obtained even without random tunneling by fine-tuning
the bare interactions in Eq.~(\ref{eq:V}) in such a way that the
neutral mode decouples. Such a fine-tuning yields $U_{-2/3,1}=\sqrt{2/3}U_{1/3,1}>0$,
assuming repulsive bare interactions. Thus, renormalization by random
tunneling is essential for obtaining an attraction out of repulsion. 

The charge sector action can be diagonalized by a hyperbolic rotation
\begin{equation}
\left(\!\begin{array}{c}
\phi_{1}\\
\phi_{\frac{-2}{3}}
\end{array}\!\right)\negmedspace=\negmedspace\left(\!\begin{array}{cc}
\cosh\chi & \sinh\chi\\
\sinh\chi & \cosh\chi
\end{array}\!\right)\negmedspace\left(\!\begin{array}{c}
\phi_{+}\\
\phi_{-}
\end{array}\!\right)\!,\,\tanh2\chi\negmedspace=\negmedspace\frac{-2U_{\frac{-2}{3},1}}{v_{\frac{2}{3}}+v_{1}}.\label{eq:rot}
\end{equation}
\emph{}Using Eq.~(\ref{eq:rot}), one finds that the scaling dimension
$\Delta$ of a generic vertex operator $O=\exp i(c_{n}\phi_{n}+c_{1}\phi_{1}+c_{2/3}\phi_{-2/3})$
is 
\begin{equation}
\Delta_{\mathbf{c}}=\frac{1}{4}(c_{1}+c_{2/3})^{2}e^{2\chi}+\frac{1}{4}(c_{1}-c_{2/3})^{2}e^{-2\chi}+\frac{1}{2}c_{n}^{2}\,.\label{eq:dim}
\end{equation}
The attractive $U_{-2/3,1}$ in Eq.~(\ref{eq:Vfp}) makes the pairing
correlation function the slowest decaying one. The superconducting
pairing correlation function in the original basis is $O_{SC}\sim e^{i(\phi_{1}-\phi_{-1})}$
{[}this operator creates two counterpropagating electrons in the $\nu=1$
modes{]}. Its dimension is calculated by first expressing $\phi_{-1}$
in terms of $\phi_{n}$ and $\phi_{-2/3}$: $e^{i(\phi_{1}-\phi_{-1})}=e^{i(\phi_{1}-\frac{1}{\sqrt{2}}[\sqrt{3}\phi_{-2/3}-\phi_{n}])}$,
and then using Eq.~(\ref{eq:dim}). We find the scaling dimension~\footnote{The pairing (charge 2) operator  $e^{i(\phi_{1}-2\phi_{-1}-3\phi_{1/3})} = e^{i(\phi_{1}-\phi_{-1})} e^{-i\sqrt{2}\phi_{n}}$ has the same scaling dimension as $O_{SC}$ while other pairing operators are less relevant~\cite{PhysRevLett.117.276804}. The existence of these equally relevant pairing operators does not change the observable effects outlined below. }
$\Delta_{SC}=\frac{1}{4}(1+\sqrt{\frac{3}{2}})^{2}e^{-2\chi}+\frac{1}{4}(1-\sqrt{\frac{3}{2}})^{2}e^{2\chi}+\frac{1}{4}$.
For $\chi\gtrsim0.26$ we have $\Delta_{SC}<1$, so the pairing correlator
decays slower than the neutral mode correlator $e^{i[3\phi_{1/3}+\phi_{-1}]}=e^{i\sqrt{2}\phi_{n}}$.
Likewise, the diagonal density operator $O_{c,\pm1}(x)\sim\partial_{x}\phi_{\pm1}$
has $\Delta=1$ irrespective of $U_{-2/3,1}$ and so the density perturbation
decays faster than pairing. Finally, we consider the off-diagonal
density operator~\citep{MIRANDA2003} $O_{CDW}\sim e^{i(\phi_{1}+\phi_{-1})}$.
We find $\Delta_{CDW}=\frac{1}{4}(1+\sqrt{\frac{3}{2}})^{2}e^{2\chi}+\frac{1}{4}(1-\sqrt{\frac{3}{2}})^{2}e^{-2\chi}+\frac{1}{4}$.
Since $\chi>0$ for $U_{-2/3,1}<0$ {[}Eq.~(\ref{eq:rot}){]}, we
always have $\Delta_{SC}<\Delta_{CDW}$. Thus, superconducting pair
correlations are the slowest decaying ones in the strong coupling
fixed point. Next, we discuss the measurable effects of this attraction. 

\begin{figure}[t] \includegraphics[width=1\columnwidth]{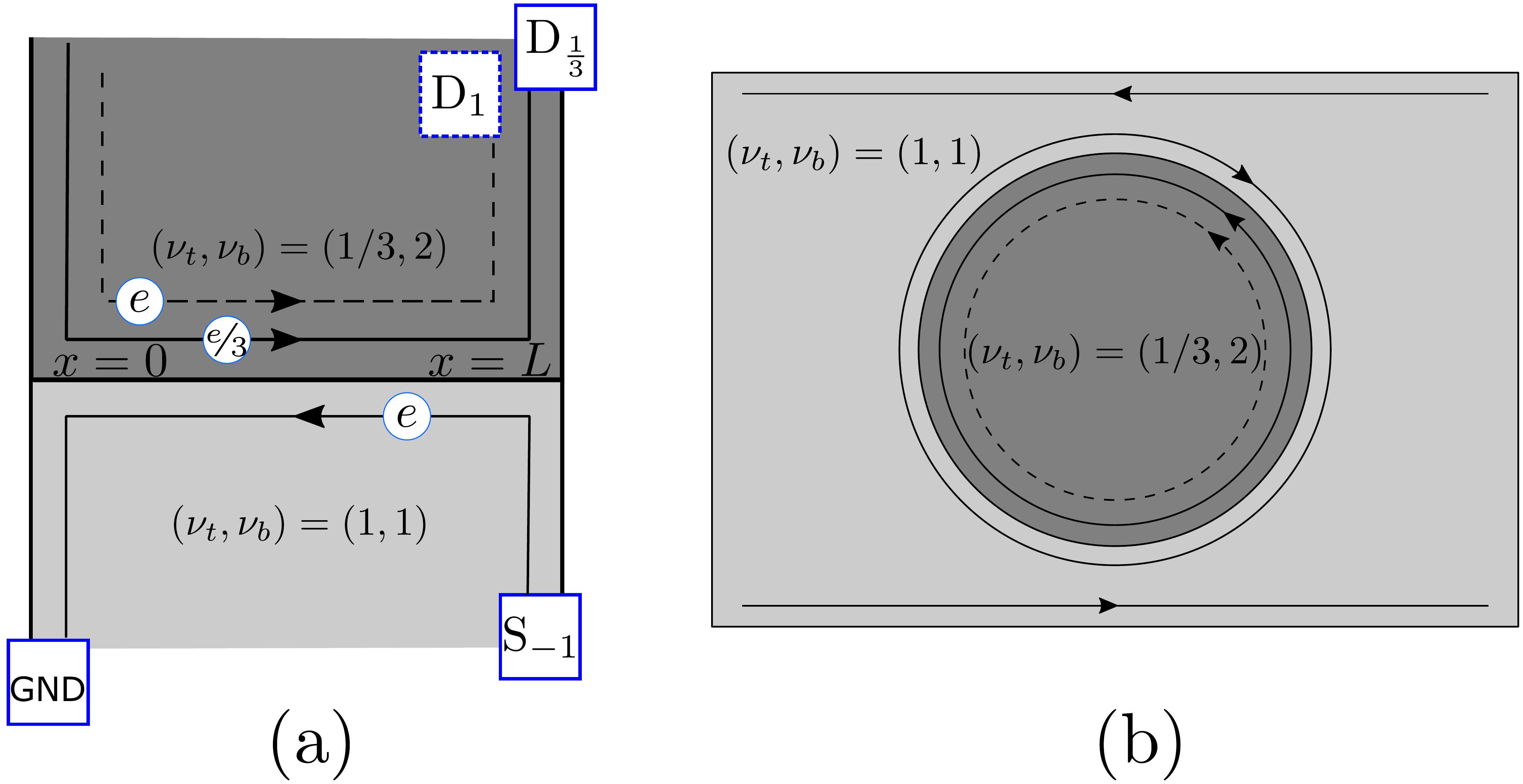}\caption{Experimental probes to measure attractive interactions in (a) shot noise, and  (b) Coulomb blockaded Hall droplet.
This configuration can be experimentally realized in a bilayer structure at an interface of bulk fillings $(\nu_{\text{top}},\nu_{\text{bottom}})=(1,1)$ and $(1/3,2)$. We have drawn  the bare modes at the interface in the same order as in Fig.~1 left. 
These modes are renormalized to an effective neutral and charge modes ($e$ and $2e/3$), see Fig. 1 right.
The dashed edge mode lives on the bottom layer and solid ones on the top. We have not drawn the $\nu=1$ mode of the bottom layer that encircles the entire system since it is not relevant to the physics at the interface. 
\label{fig:2}} \end{figure}

\textbf{\emph{Consequences of attraction.}} \emph{}The relatively
long-ranged pairing correlations are a direct consequence of the attractive
interaction $U_{-2/3,1}<0$ in Eq.~(\ref{eq:Vfp}). Thus, one way
to probe our proposed fixed point is to measure $U_{-2/3,1}$ or its
sign. Since the fixed point  action is that of a non-chiral spinless
Luttinger liquid, one is faced with the known task of measurement
of the Luttinger liquid parameter. Next, we outline three possible
ways to do this. We focus on the experimentally relevant bilayer quantum
Hall system, see Fig.~\ref{fig:2}.

\emph{Signature of attraction in shot noise.} It is well-known that
the interaction parameter in a non-chiral Luttinger liquid can be
measured with a.c. shot noise~\citep{PhysRevLett.92.226405,Berg09,Kuehne15}.
The attractive interactions in our setup can be measured in a similar
experiment, see Fig.~\ref{fig:2}a. Employing the theory of inhomogeneous
Luttinger liquid~\citep{PhysRevB.52.R17040}, we solve~\cite{Note2}
the problem of a bare incoming mode $\phi_{-1}$ scattered off an
interacting region at the superconducting fixed point. In particular,
the charge reflected into the mode $\phi_{1}$ (drain $D_{1}$ in
the bottom layer) is fractional with a non-universal magnitude. Its
sign however is given directly by $U_{-2/3,1}$. \emph{Thus, a smoking
gun signature of the emergence of the attraction would be negative
current measured at} $D_{1}$  (``Andreev reflection'' at the edge)~\footnote{Compare with the emergence of negative currents in Ref.~\citep{2017AnPhy.385..287P}.}.
The reflected charge can be measured in a time-domain experiment and
requires access to frequencies $\omega\gtrsim v/L$ where $v=\max(v_{2/3},v_{1})$
and $L$ is the length of the scattering region. 

\emph{Signature of attraction in tunneling conductance.} One can also
measure $U_{-2/3,1}$ from tunneling conductance~\citep{wen2004quantum,PhysRevLett.109.026803,PhysRevB.90.075403,2017NatPh..13..491S,2018arXiv181006871P,PhysRevLett.117.276804} in the interacting
region, for example by using a point-contact to an auxiliary $\nu=1$
edge. For describing the tunneling Hamiltonian, consider the vertex
operator $e^{i(n_{1}\phi_{1}-n_{-1}\phi_{-1}+3n_{1/3}\phi_{1/3})}$
that creates an excitation of total charge $n_{1}+n_{-1}+n_{1/3}$
on the edge; here $n_{1},n_{-1},n_{1/3}\in\mathbb{Z}$. The contribution
to the tunneling current from the above operator exhibits a power-law
bias voltage dependence~\citep{wen2004quantum},~\footnote{The result is valid at low temperatures $k_B T \ll eV$. At higher temperatures the current exhibits a power-law in temperature, $I\propto V T^{2\alpha-2}$. } $I\propto V^{2\alpha-1}$
where the exponent $\alpha=\Delta_{n_{1},n_{-1},n_{1/3}} +\frac{1}{2} n$ is determined by the scaling dimension $\Delta_{n_{1},n_{-1},n_{1/3}}$
{[}obtained from Eq.~(\ref{eq:dim}) after transforming the vertex
operator into the neutral mode basis by using Eq.~(\ref{eq:neutralcharge}){]} 
and the number of electrons $n$ removed from the auxiliary edge.
The total tunneling current is a sum of elementary tunneling processes,
but will be dominated at small voltages by those with a low value
of $\alpha$. For moderate interaction strengths $\chi$ {[}Eq.~(\ref{eq:rot}){]}
the dominant contributions are the 1-electron tunneling operators
$e^{i\phi_{1}}$ and $e^{-i\phi_{-1}}$, as well as the 2-electron
tunneling operator $e^{i(\phi_{1}-\phi_{-1})}$. Their respective
tunneling amplitudes $t_{1}$, $t_{-1}$, and $t_{1}t_{-1}$, are
in principle controllable by gating, so that different 1-electron
contributions can be turned on and off. The signature of attractive
interactions ($\chi>0$) is that $\Delta_{1,-1,0}<\Delta_{1,0,0}+\Delta_{0,-1,0}$,
meaning that when tunneling to both $\nu=1$ edges is present, the
current is less suppressed by a small bias than one would expect from
uncorrelated tunnelings to each edge separately.

\emph{Signature of attraction in a mesoscopic droplet.} Finally, one
can perform a fully thermodynamic measurement in a Coulomb blockaded
quantum Hall droplet, see Fig.~\ref{fig:2}b. This is akin to ideas
of ``attraction from repulsion'' that have been implemented in other
systems~\citep{2016Natur.535..395H}, compare also proposals to probe
neutral modes in the context of quantum Hall edges~\citep{PhysRevLett.114.156401}.
The signature of attraction in the Coulomb blockaded droplet is $2e$-periodic
charge transitions as a gate charge is varied~\cite{Note2}. This
signature can be measured in a thermodynamic capacitive measurement
of the charge or in a transport measurement of the Coulomb peak spacings.

\textbf{\emph{Discussion.}} Our proposal relies on the tunneling between
the modes $\phi_{1/3}$ and $\phi_{-1}$ being the most RG relevant
perturbation. Typically, the tunneling $e^{i(\phi_{1}+\phi_{-1})}$
between the $\nu=1$ modes is also relevant and leads to a trivial
localization of the $\nu=1$ modes. This effect should however be
present only at very low temperatures since we expect the bare amplitude
of the $\nu=1$ tunneling to be very weak due to the large separation
of the $\nu=1$ modes. The $\nu=1$ tunneling can also be entirely
avoided by considering a setup with spin-polarized Landau levels where
$\phi_{1}$ and $\phi_{-1}$ have opposite spins and the tunneling
between them is forbidden by spin conservation. This can be achieved
with an interface between $(\nu_{\text{top}},\nu_{\text{bottom}})=(1/3,2)$
and $(\nu_{\text{top}},\nu_{\text{bottom}})=(1,1)$, assuming the
$\nu=2$ state consists of opposite-polarized $\nu=1$ states. This
state also satisfies the requirement that $\nu_{\text{bottom}}\geq\nu_{\text{top}}$
holds on both sides~\citep{2018NatPh..14..411R}.

In our model we assumed that $U_{1/3,-1}$ is the largest interaction
while the other two were treated perturbatively, which ensures that
$e^{i[3\phi_{1/3}+\phi_{-1}]}$ is relevant and KFP fixed point is
reached.  Thus, we rely on the double-inequality $U_{1/3,-1}>U_{1/3,1}>3U_{-1,1}$
to approach the fixed point with attractive interactions.  We find
that Coulomb interaction screened by a nearby gate electrode~\citep{PhysRevLett.77.1354},\cite{Note2}
allows both inequalities to be satisfied. 

One may ask how essential the bilayer construction is to manifest
our theory. For example, edge reconstruction in a $\nu=1/n$ Laughlin
state can give rise to a $\nu=1/m$ stripe in the bulk-vacuum interface.
Disordered tunneling between the two inner modes gives rise to counterpropagating
neutral and a charge modes. The propagation directions of these modes
are determined by comparing the two filling fractions. If $n<m$,
the charge mode is co-propagating with the outermost $\nu=1/m$ mode.
Therefore, there are no emerging superconducting correlations even
if there is attraction between the two charge modes. (Interactions
between co-propagating modes do not affect the scaling dimensions
of the operators involved, since the $V$-matrix can be diagonalized
with an orthogonal transformation~\citep{wen2004quantum}.) In the
more interesting scenario $n>m$, the charge modes are counterpropagating
and superconducting correlations may in principle emerge. In this
case the  the interaction is attractive when $U_{\frac{1}{n},\frac{1}{m}}>\frac{n}{m}U_{-\frac{1}{m},\frac{1}{m}}$.
However, for an interaction falling monotonically with distance, we
expect $U_{\frac{1}{n},\frac{1}{m}}<U_{-\frac{1}{m},\frac{1}{m}}$
because the outermost mode $+1/m$ is closer to $-1/m$ rather than
the bulk mode $1/n$. This is why we do not expect to find superconducting
correlations in such a simple model of edge reconstruction. This problem
is circumvented in the bilayer setup, see Fig.~\ref{fig:1}. 
Here the two-dimensional electron gas (2DEG) is replaced by a bilayer of 2DEGs whose individual filling fractions can be tuned. 
The resulting boundary consists of chiral mode structure which can be controlled on-demand by tuning back gate voltages and the magnetic field.
Finally,
we note that our proposal also works for an interface between $(\nu_{\text{top}},\nu_{\text{bottom}})=(2/3,1)$
and $(\nu_{\text{top}},\nu_{\text{bottom}})=(0,2)$, assuming that
the $\nu=2/3$ edge consists of counterpropagating $\nu=1$ and $\nu=1/3$
modes~\citep{1990PhRvL..64..220M,PhysRevLett.64.2206}.
\begin{acknowledgments}
We thank M. Heiblum, R. Lutchyn, and D. Pikulin for discussions. J.I.V.
thanks the Aspen Center for Physics which is supported by National
Science Foundation grant PHY-1607611. M.G. was supported by the Israel
Science Foundation (Grant No. 227/15), the German Israeli Foundation
(Grant No. I-1259-303.10), the US-Israel Binational Science Foundation
(Grant No. 2016224), and the Israel Ministry of Science and Technology
(Contract No. 3-12419). Y.G. was supported by DFG RO 2247/11-1 and
CRC 183 (project C01), and the Italia-Israel project QUANTRA.
\end{acknowledgments}

\bibliographystyle{apsrev4-1}
\bibliography{refs}



\newpage
\begin{widetext}

\section*{Supplementary Material to ``Superconducting correlations out of repulsive
interactions on a fractional quantum Hall edge''}

In this Supplementary Material, we present the full RG equations at
the fine-tuned point $U_{n,-2/3}=0$, discuss in more detail the signatures
of attraction in the shot noise and Coulomb blockaded droplet, and
finally outline the geometrical requirements to find attraction from
repulsion in a bilayer system.

\subsection{Renormalization group flow of the $V$-matrix} 

In this section we study the fine-tuned KFP fixed point which corresponds
to a bare interaction $U_{1/3,-1}=3(v_{1/3}+v_{-1})/4$ which means
$\Delta_{3,1,0}=1$. This fine-tuned point is easy to study because
the neutral mode is automatically decoupled from $\phi_{-2/3}$, since
in Eq.~(\ref{eq:Vn}) of the main text we have $U_{n,-2/3}=0$: 
\begin{equation}
\mathbf{V}=\left(\begin{array}{ccc}
v_{n} & 0 & U_{n1}\\
0 & v_{2/3} & U_{-2/3,1}\\
U_{n1} & U_{-2/3,1} & v_{1}
\end{array}\right)_{(n)}\,.\label{eq:VnSM}
\end{equation}
For completeness, the non-zero interactions in the original basis
are $U_{n1}=\frac{1}{\sqrt{2}}(U_{\frac{1}{3},1}-U_{-1,1})$ and $U_{-2/3,1}=\frac{1}{\sqrt{6}}(3U_{-1,1}-U_{\frac{1}{3},1})$.
Treating the interactions $U_{n1}$,~$U_{-2/3,1}$ perturbatively,
we can diagonalize $\mathbf{V}$ with a transformation $M$ that preserves
$\mathbf{K}=\text{diag}(1,-1,1)_{(n)}$:
\begin{equation}
\mathbf{V}^{(D)}=M\mathbf{V}M^{T},\,M=\left(\begin{array}{ccc}
1 & 0 & -a_{1}\\
0 & 1 & a_{2}\\
a_{1} & a_{2} & 1
\end{array}\right)_{(n)}\,,
\end{equation}
where $a_{1}=\frac{U_{n1}}{v_{n}-v_{1}}$, $a_{2}=\frac{U_{-2/3,1}}{v_{2/3}+v_{1}}$.
The tunneling vector $\mathbf{c}=(\sqrt{2},0,0)_{(n)}$ transforms
to $\mathbf{c}^{(D)T}=M^{T}(\sqrt{2},0,0)_{(n)}^{T}=\sqrt{2}(1,0,-a_{1})_{(n)}^{T}$. 

Next, we follow Ref.~\citep{PhysRevB.66.115305} to find how $\mathbf{V}$
flows upon renormalization. To first order in $U_{n1}$,~$U_{-2/3,1}$,
the flow is entirely due to $U_{n1}$ which couples to disordered
neutral mode. The RG equation is 
\begin{equation}
\frac{d\mathbf{V}^{(D)}}{dl}=\left(\begin{array}{ccc}
0 & 0 & (v_{n}-v_{1})\frac{d\theta}{dl}\\
0 & 0 & 0\\
(v_{n}-v_{1})\frac{d\theta}{dl} & 0 & 0
\end{array}\right)_{(n)}\,,
\end{equation}
where 
\begin{flalign}
(v_{n}-v_{1})\frac{d\theta}{dl} & =4\pi\frac{2a_{1}v_{n}v_{1}}{v_{n}^{2}v_{1}^{2a_{1}^{2}}}W\approx8\pi\frac{U_{n1}}{v_{n}-v_{1}}\frac{v_{1}}{v_{n}}W\,.
\end{flalign}
This corresponds to 
\begin{equation}
\frac{dU_{n1}}{dl}=-8\pi\frac{1}{v_{1}-v_{n}}\frac{v_{1}}{v_{n}}U_{n1}W\,.\label{smeq:Un1}
\end{equation}
Ignoring the term $U_{-2/3,1}$ in Eq.~(\ref{eq:VnSM}), the action
corresponds to a neutral mode coupled to a co-moving charge mode.
This system has a stable fixed point~\citep{PhysRevB.66.115305}
$U_{n1}\to0$ when $v_{1}>v_{n}$. 

To find beyond tree-level accuracy, we can diagonalize $\mathbf{V}^{(n)}$
working to 2nd order accuracy. The tunneling vector is 
\begin{equation}
\mathbf{c}^{(D)}=\sqrt{2}\left(1-\frac{U_{n1}^{2}}{2\left(v_{1}-v_{n}\right){}^{2}},\,-\frac{U_{-2/3,1}U_{n1}}{\left(v_{n}-v_{1}\right)\left(v_{2/3}+v_{n}\right)},\,\frac{U_{n1}}{v_{n}-v_{1}}\right)_{(n)}^{T}\,,
\end{equation}
which shows that $\frac{dU_{-2/3,1}}{dl}$ starts at order $\sim U_{-2/3,1}U_{n1}^{2}$.
One therefore has to go to 3rd order to capture this accurately. Such
a calculation gives
\begin{equation}
\frac{dU_{-2/3,1}}{dl}=-8\pi\frac{U_{-2/3,1}U_{n1}^{2}}{\left(v_{n}-v_{1}\right)^{2}\left(v_{2/3}+v_{n}\right)}\frac{v_{1}v_{2/3}W}{v_{n}^{2}}\,.\label{smeq:U23}
\end{equation}
Thus $U_{-2/3,1}$ is irrelevant but very marginally so: only to second
order in $U_{n1}^{2}$. The 3rd order corrections to Eq.~(\ref{smeq:Un1})
modify its RHS by an overall factor $1+\frac{U_{-2/3,1}^{2}}{(v_{1}-v_{n})(v_{2/3}+v_{n})}-\frac{2U_{n1}^{2}}{(v_{1}-v_{n})^{2}}$.
This correction is inconsequential at small interaction strengths. Solving
Eqs.~(\ref{smeq:Un1}) {[}including the aforementioned correction{]}
and~(\ref{smeq:U23}) numerically and expressing them in the original
basis leads to Fig.~2 of the main text.

\subsection{Signature of attraction in shot noise}

We focus on a geometry where the modes $\phi_{-1},\,\phi_{1/3}$ are
uncoupled at $x\to\pm\infty$. In the scattering region $0<x<L$ the
modes couple as described by the bare action~(\ref{eq:ShomogDiab}),
see Fig.~\ref{fig:2}a. We assume that the fixed point of a decoupled
neutral mode is reached throughout the entire scattering region. Let
us consider an incoming mode $\phi_{-1}$ from $x=+\infty$, which
gets reflected at $x=L$ into modes $\phi_{1/3},\,\phi_{1}$ upon
encountering the scattering region. For $x>L$, the eigenmodes of
the system are $\phi_{-1},\,\phi_{1},\,\phi_{1/3}$. They are described
by an action {[}$\overline{\phi}_{1/3}=\sqrt{3}\phi_{1/3}${]} 

\begin{equation}
S=\frac{1}{4\pi}\int d\tau\int_{L}^{\infty}dx\left[\partial_{x}\phi_{1}i\partial_{\tau}\phi_{1}+\partial_{x}\overline{\phi}_{1/3}i\partial_{\tau}\overline{\phi}_{1/3}-\partial_{x}\phi_{-1}i\partial_{\tau}\phi_{-1}+v_{1}(\partial_{x}\phi_{1})^{2}+v_{1/3}(\partial_{x}\overline{\phi}_{1/3})^{2}+v_{-1}(\partial_{x}\phi_{-1})^{2}\right]\,.
\end{equation}
For $x<L$, we have first the action of decoupled neutral mode and
coupled charge modes, 
\begin{flalign}
S & =\frac{1}{4\pi}\int d\tau\int_{-\infty}^{L}dx\left[\partial_{x}\phi_{n}i\partial_{\tau}\phi_{n}+\partial_{x}\phi_{1}i\partial_{\tau}\phi_{1}-\partial_{x}\phi_{-2/3}i\partial_{\tau}\phi_{-2/3}\right.\nonumber \\
 & +\left.v_{n}(\partial_{x}\phi_{n})^{2}+v_{2/3}(\partial_{x}\phi_{-2/3})^{2}+v_{1}(\partial_{x}\phi_{1})^{2}+2U_{-2/3,1}\partial_{x}\phi_{-2/3}\partial_{x}\phi_{1}\right]\,,\label{smeq:ScSn}
\end{flalign}
where $\phi_{n}=\frac{1}{\sqrt{2}}\left(\sqrt{3}\overline{\phi}_{1/3}+\phi_{-1}\right)$
and $\phi_{-2/3}=\sqrt{\frac{3}{2}}\left(\frac{1}{\sqrt{3}}\overline{\phi}_{1/3}+\phi_{-1}\right)$.
Here, we can diagonalize the charge sector by 
\begin{equation}
\left(\begin{array}{c}
\phi_{1}\\
\phi_{-2/3}
\end{array}\right)=\left(\begin{array}{cc}
\cosh\chi & \sinh\chi\\
\sinh\chi & \cosh\chi
\end{array}\right)\left(\begin{array}{c}
\phi_{+}\\
\phi_{-}
\end{array}\right)\,,\quad\tanh2\chi=-\frac{2U_{-2/3,1}}{v_{2/3}+v_{1}}\,,
\end{equation}
to get the diagonal action, 
\begin{equation}
S=\frac{1}{4\pi}\int d\tau\int_{-\infty}^{L}dx\left[\partial_{x}\phi_{n}i\partial_{\tau}\phi_{n}+\partial_{x}\phi_{+}i\partial_{\tau}\phi_{+}-\partial_{x}\phi_{-}i\partial_{\tau}\phi_{-}+v_{n}(\partial_{x}\phi_{n})^{2}+v_{+}(\partial_{x}\phi_{+})^{2}+v_{-}(\partial_{x}\phi_{-})^{2}\right]\,.
\end{equation}
The eigenbases at $x<L$ and $x>L$ are related by 
\begin{equation}
\left(\begin{array}{c}
\phi_{1}\\
\overline{\phi}_{1/3}\\
\phi_{-1}
\end{array}\right)=\left(\begin{array}{ccc}
0 & \sqrt{\frac{3}{2}} & \frac{1}{\sqrt{2}}\\
1 & 0 & 0\\
0 & \frac{1}{\sqrt{2}} & \sqrt{\frac{3}{2}}
\end{array}\right)^{-1}\left(\begin{array}{c}
\phi_{n}\\
\phi_{1}\\
\phi_{-2/3}
\end{array}\right)=\left(\begin{array}{ccc}
0 & \cosh\chi & \sinh\chi\\
\frac{\sqrt{6}}{3\sqrt{3}-1} & \frac{\sqrt{2}\sinh\chi}{1-3\sqrt{3}} & \frac{\sqrt{2}\cosh\chi}{1-3\sqrt{3}}\\
\frac{\sqrt{2}}{1-3\sqrt{3}} & \frac{3}{26}\left(\sqrt{2}+3\sqrt{6}\right)\sinh\chi & \frac{3}{26}\left(\sqrt{2}+3\sqrt{6}\right)\cosh\chi
\end{array}\right)\left(\begin{array}{c}
\phi_{n}\\
\phi_{+}\\
\phi_{-}
\end{array}\right)\,.
\end{equation}

Let us next consider the following scattering problem. We suppose
that there is an incoming wavepacket in the mode $\phi_{-1}(x,t)=\Theta(v_{-1}t+[x-L])$
coming from $+\infty$, and that it will enter the coupled region
at $t=0$. The reflected waves will be $\overline{\phi}_{1/3}(x,t)=R_{1/3}\Theta(v_{1/3}t-[x-L]-B_{1/3})$
and $\phi_{1}(x,t)=R_{1}\Theta(v_{1}t-[x-L]-B_{1})$ in the region
$x>L$. Here $B_{i}>0$ and $R_{i}$ are unknown. This means that
there will be charges $R_{1}$ and $R_{1/3}$ reflected into the respective
drains. For $x<L$ we have wave solutions $f_{j}(v_{j}t-\sigma_{j}[x-L])$
for $j=n,\pm$ {[}$\sigma_{j}=-1$ for $\phi_{-}$ and $+1$ for the
other two modes{]}. Assuming there is no incoming wave from the left,
we have that $f_{j}(x,t=0)=0$ for all $j$, $x<L$. From this it
follows that $f_{+}=f_{n}=0$ identically. Likewise, $f_{-}(z)=0$
for $z<0$. Finally, continuity of the waves at $x=L$ gives the relation
\begin{equation}
\left(\begin{array}{c}
R_{1}\Theta(v_{1}t-B_{1})\\
R_{1/3}\Theta(v_{1/3}t-B_{1/3})\\
\Theta(v_{-1}t)
\end{array}\right)=\left(\begin{array}{ccc}
0 & \cosh\chi & \sinh\chi\\
\frac{\sqrt{6}}{3\sqrt{3}-1} & \frac{\sqrt{2}\sinh\chi}{1-3\sqrt{3}} & \frac{\sqrt{2}\cosh\chi}{1-3\sqrt{3}}\\
\frac{\sqrt{2}}{1-3\sqrt{3}} & \frac{3}{26}\left(\sqrt{2}+3\sqrt{6}\right)\sinh\chi & \frac{3}{26}\left(\sqrt{2}+3\sqrt{6}\right)\cosh\chi
\end{array}\right)\left(\begin{array}{c}
0\\
0\\
f_{-}(v_{-}t)
\end{array}\right)\,,
\end{equation}
or 
\begin{equation}
\left(\begin{array}{c}
R_{1}\Theta(v_{1}t-B_{1})\\
R_{1/3}\Theta(v_{1/3}t-B_{1/3})\\
\Theta(v_{-1}t)
\end{array}\right)=\left(\begin{array}{c}
\sinh\chi\\
\frac{\sqrt{2}}{1-3\sqrt{3}}\cosh\chi\\
\frac{3}{26}\left(\sqrt{2}+3\sqrt{6}\right)\cosh\chi
\end{array}\right)f_{-}(v_{-}t)\,.
\end{equation}
From the last equation we find $f_{-}(z)=\Theta(\frac{v_{-1}}{v_{-}}z)/\frac{3}{26}\left(\sqrt{2}+3\sqrt{6}\right)\cosh\chi$
for $z>0$. Note that $\frac{3}{26}\left(\sqrt{2}+3\sqrt{6}\right)=\frac{3\sqrt{2}}{\left(3\sqrt{3}-1\right)}$.
Thus, the function $f_{-}$ is determined fully to be $f_{-}(z)=\Theta(\frac{v_{-1}}{v_{-}}z)\left(3\sqrt{3}-1\right)/3\sqrt{2}\cosh\chi$.
This is the transmitted wave. The remaining two equations yield $B_{1}=B_{1/3}=0$
and reflection coefficients 
\begin{equation}
R_{1}=\frac{\left(3\sqrt{3}-1\right)\tanh\chi}{3\sqrt{2}}\approx-1.011\frac{1-g}{1+g}\,,\quad R_{1/3}=-\frac{1}{3}\,,
\end{equation}
where $\frac{\left(3\sqrt{3}-1\right)}{3\sqrt{2}}\approx1.011$ and
we introduced the ``Luttinger liquid parameter'' $g=\sqrt{\frac{\frac{1}{2}(v_{2/3}+v_{1})-U_{-2/3,1}}{\frac{1}{2}(v_{2/3}+v_{1})+U_{-2/3,1}}}$.
The charges reflected into the drains $1$ and $1/3$ are respectively
$R_{1}$ and $R_{1/3}$.  When $U_{-2/3,1}>0$ {[}repulsive interaction{]},
we have $g<1$ and correspondingly $-1.011\lesssim R_{1}<0$. On the
other hand, when $U_{-2/3,1}<0$ [attraction], we have $g>1$ and $0<R_{1}\lesssim1.011$.
Therefore, the crucial signature of attraction is the sign of $R_{1}$. It can be measured by measuring the current in the $\nu =1$ edge in time-domain.  

\subsection{Signature of attraction in a mesoscopic droplet}

In a periodic finite system of length $L$, we have the mode expansion
{[}$\nu=\pm1,\,1/3${]}
\begin{equation}
\phi_{\nu}(x)=\frac{2\pi\nu x}{L}N_{\nu}+\varphi_{\nu}+\delta\phi_{\nu}(x)\,,
\end{equation}
where $[\varphi_{\nu},N_{\nu}]=i$ and $\varphi_{\nu},N_{\nu}$ commute
with $\delta\phi_{\nu}$. Using the mode expansion, we find for the
charge sector of the fixed point action {[}obtained from Eq.~(5)
of the main text, see also Eq.~(\ref{smeq:ScSn}){]}
\begin{flalign}
S_{\text{charge}} & =\int d\tau dx\frac{1}{4\pi}\left[\partial_{x}\delta\phi_{1}i\partial_{\tau}\delta\phi_{1}-\partial_{x}\delta\phi_{-2/3}i\partial_{\tau}\delta\phi_{-2/3}\right]\\
 & +\int d\tau dx\frac{1}{4\pi}\left[v_{2/3}(\partial_{x}\delta\phi_{-2/3})^{2}+v_{1}(\partial_{x}\delta\phi_{1})^{2}+2U_{-2/3,1}\partial_{x}\delta\phi_{-2/3}\partial_{x}\delta\phi_{1}\right]\\
 & +\frac{1}{2L}\int d\tau dx\left[N_{1}i\partial_{\tau}\varphi_{1}-v_{2/3}N_{-2/3}i\partial_{\tau}\varphi_{c}\right]\\
 & +\frac{\pi}{L}\int d\tau\left[v_{2/3}N_{-2/3}^{2}+v_{1}N_{1}^{2}+2U_{-2/3,1}N_{-2/3}N_{1}\right]\,,
\end{flalign}
where 
\begin{equation}
N_{-2/3}=\sqrt{\frac{3}{2}}\left(\frac{1}{3}N_{1/3}+N_{-1}\right)\,,\quad\varphi_{c}=\sqrt{\frac{3}{2}}\left(\varphi_{1/3}+\varphi_{-1}\right)\,.
\end{equation}
 For the neutral sector we have {[}Eq.~(5) of the main text, Eq.~(\ref{smeq:ScSn}){]}
\begin{equation}
S_{\text{neutral}}=\int d\tau dx\frac{1}{4\pi}\left[\partial_{x}\delta\phi_{n}i\partial_{\tau}\delta\phi_{n}+v_{n}(\partial_{x}\delta\phi_{n})^{2}+\frac{2\pi}{L}N_{n}i\partial_{\tau}\varphi_{n}+v_{n}[\frac{2\pi}{L}N_{n}]^{2}\right]\,,
\end{equation}
where $N_{n}=\sqrt{\frac{1}{2}}\left(N_{1/3}+N_{-1}\right)$ and $\varphi_{n}=\sqrt{\frac{1}{2}}\left(3\varphi_{1/3}+\varphi_{-1}\right)$. 

The ``charging'' Hamiltonian obtained from above is 
\begin{equation}
H_{c}=\frac{\pi}{L}\left[v_{n}N_{n}^{2}+v_{2/3}N_{-2/3}^{2}+v_{1}N_{1}^{2}+2U_{-2/3,1}N_{-2/3}N_{1}\right]\,.
\end{equation}
The stability of $\mathbf{V}_{\text{f.p.}}$ requires that $\sqrt{v_{2/3}v_{1}}>|U_{-2/3,1}|$.
This ensures that $H_{c}$ is positive, 
\begin{equation}
v_{2/3}N_{-2/3}^{2}+v_{1}N_{1}^{2}+2U_{-2/3,1}N_{-2/3}N_{1}>\left(\sqrt{v_{2/3}}N_{-2/3}-\sqrt{v_{1}}N_{1}\right)^{2}\,.
\end{equation}
In reality, the charging Hamiltonian is dominated by the total charging
energy, which arises from the long-range Coulomb interaction. The
total charge is given by $N_{\text{tot}}=N_{1}+N_{-1}+\frac{1}{3}N_{1/3}=N_{1}+\sqrt{\frac{2}{3}}N_{-2/3}$,
and the charging energy is then 
\begin{equation}
E_{c}(N_{1}+\sqrt{\frac{2}{3}}N_{-2/3}-n_{g})^{2}\,,
\end{equation}
where $en_{g}$ is the controllable induced gate charge. Let us see
how the attractive interaction $\frac{\pi}{L}U_{-2/3,1}N_{-2/3}N_{1}$
affects the $n_{g}$-dependence of the ground state charge. Note that
$N_{1}$ is integer while $N_{-2/3}=\sqrt{3/2}\times\text{integer}.$
The total energy is 
\begin{equation}
E_{c}(N_{1}+\sqrt{\frac{2}{3}}N_{-2/3}-n_{g})^{2}+\frac{\pi}{L}\left[v_{2/3}N_{-2/3}^{2}+v_{1}N_{1}^{2}+2U_{-2/3,1}N_{-2/3}N_{1}\right]\,.
\end{equation}
We include the terms $v_{2/3},\,v_{1}$ to ensure bounded spectrum.
In our model $L$ is the system length. However, our bosonization
does not treat accurately the long-range Coulomb interaction so we
cannot obtain quantitative estimates. The qualitative findings outlined
below should however remain true.

For simplicity, let us focus on the four states $(N_{1},\sqrt{\frac{2}{3}}N_{-2/3})=\{(0,0),\,(1,0),\,(0,1),\,(1,1)\}$.
Relative to the $(0,0)$ state, the other have energies 
\begin{flalign}
E(1,0) & =E_{c}(1-2n_{g})+\frac{\pi}{L}v_{1}\,,\\
E(0,1) & =E_{c}(1-2n_{g})+\frac{\pi}{L}v_{2/3}\,,\\
E(1,1) & =4E_{c}(1-n_{g})+\frac{\pi}{L}\left[v_{2/3}+v_{1}+2U_{-2/3,1}\right]\,.
\end{flalign}
We have $E(1,1)=0$ when 
\begin{equation}
E_{c}(1-2n_{g})=-\left(E_{c}+\frac{\pi}{2L}\left[v_{2/3}+v_{1}+2U_{-2/3,1}\right]\right)\,.
\end{equation}
For this value of $n_{g}$, the other two energies are 
\begin{flalign}
E(1,0) & =-\frac{\pi}{L}U_{-2/3,1}-\left(E_{c}+\frac{\pi}{2L}\left[v_{2/3}-v_{1}\right]\right)\,,\\
E(0,1) & =-\frac{\pi}{L}U_{-2/3,1}-\left(E_{c}+\frac{\pi}{2L}\left[-v_{2/3}+v_{1}\right]\right)\,,
\end{flalign}
which are positive when 
\begin{equation}
-U_{-2/3,1}>\frac{L}{\pi}E_{c}\pm\frac{1}{2}\left[v_{2/3}-v_{1}\right]\,.
\end{equation}
Under this condition we have a direct transition from $(0,0)$ ground
state to $(1,1)$ ground state as $n_{g}$ is tuned. We have an earlier
condition $\sqrt{v_{2/3}v_{1}}>|U_{-2/3,1}|$ from stability. This
imposes the constraint 
\begin{equation}
2\sqrt{v_{2/3}v_{1}}>2\frac{L}{\pi}E_{c}\pm\left[v_{2/3}-v_{1}\right]\,.
\end{equation}
This is a condition on $E_{c}$. For example, when $v_{2/3}>v_{1}$,
the signature $(0,0)\to(1,1)$ transition exists when $2\frac{L}{\pi}E_{c}<2\sqrt{v_{2/3}v_{1}}-\left[v_{2/3}-v_{1}\right]=2v_{1}-(\sqrt{v_{2/3}}-\sqrt{v_{1}})^{2}$.

\subsection{Geometric requirements for the bilayer in the case of long-range
Coulomb repulsion}

In the main text we found the requirement $3U_{-1,1}<U_{1/3,1}$ in
order to get attraction, $U_{-2/3,1}<0$, between the charge modes
at the disordered fixed point. In order to ensure that we flow to
the disordered fixed point, we further require $U_{\frac{1}{3},-1}>U_{\frac{1}{3},1}$.
In this Section, we consider Coulomb interaction $U(|\mathbf{r}|)$
and find the requirements for the bilayer geometry. (Note however
that our bosonization description assumes short-range interactions;
our estimates in this Section are therefore mostly qualitative.) We
assume that the bilayers are separated by a distance $z$, while in-plane
the modes $\phi_{-1},\,\phi_{1/3},\,\phi_{1}$ are at positions $0,y_{0},y_{0}+a$
. Thus, we have the double inequality {[}recall that $\phi_{1}$ is
separated by additional distance $z$ in the perpendicular direction{]}
\begin{equation}
U(y_{0})>U(\sqrt{a^{2}+z^{2}})>3U(\sqrt{(y_{0}+a)^{2}+z^{2}})\,.
\end{equation}
This inequality cannot be satisfied for simple Coulomb interaction
$U(|\mathbf{r}|)\sim1/|\mathbf{r}|$. However, for a faster decaying
interaction, $U(|\mathbf{r}|)\sim1/|\mathbf{r}|^{3}$, it can be satisfied~\citep{PhysRevLett.77.1354}.
(For example, when $z\to0$, we find $a>y_{0}>(\sqrt[3]{3}-1)a\approx0.44a$.)
The Coulomb interaction decays cubically when it is screened by an
external gate. Let us consider a gate planar with the bilayer at a
distance $d$ from the bottom quantum well.  If we have a bottom
gate {[}$d<z<0${]} and the mode $1$ lives in the bottom layer, we
have for example $U_{-1,1}\propto\frac{1}{\sqrt{(y_{0}+a)^{2}+z^{2}}}-\frac{1}{\sqrt{(y_{0}+a)^{2}+(2d-z)^{2}}}$.
Supposing that $y_{0},\,a\gg|d|,|z|$ we have then for example $U_{-1,1}\propto\frac{1}{2}\frac{(2d-z)^{2}-z^{2}}{(y_{0}+a)^{3}}$
which decays cubically, as promised. The inequalities in this case
become 
\begin{equation}
U_{\frac{1}{3},-1}\approx\frac{2d^{2}}{y_{0}^{3}}>U_{\frac{1}{3},1}\approx\frac{1}{2}\frac{(2d-z)^{2}-z^{2}}{a^{3}}>3U_{-1,1}\approx3\frac{1}{2}\frac{(2d-z)^{2}-z^{2}}{(y_{0}+a)^{3}}\,,
\end{equation}
or
\begin{equation}
\sqrt[3]{\frac{1}{1-\frac{z}{d}}}>\frac{y_{0}}{a}>\sqrt[3]{3}-1\,.
\end{equation}
The right inequality corresponds to the condition $3U_{-1,1}<U_{1/3,1}$
for attraction. The left one ensures that the disordered fixed point
should be reachable. The RHS is less than one, $\sqrt[3]{3}-1\approx0.44$.
On the other hand, the LHS is always larger than one for a bottom
gate, $d<z<0$. For a top gate, $z<0<d$, there is an upper bound
$d<\frac{|z|}{\frac{1}{(\sqrt[3]{3}-1)^{3}}-1}\approx0.09|z|$. 

\end{widetext}
\end{document}